\documentclass[aps,prb,reprint,groupedaddress,showpacs,floatfix]{revtex4-1}
\usepackage{graphicx}
\usepackage{dcolumn}

\begin{document}

\title{Short-time Monte Carlo simulation of the 
majority-vote model on cubic lattices}

\author{K. P. do Nascimento}
\affiliation{Departamento de Matem\'atica, Universidade
 Regional do Cariri,
 Av. Le\~ao
Sampaio, Tri\^angulo  Juazeiro do Norte, 63010-970, Brazil}
\affiliation{Departamento de Estatast\'ica e Inform\'atica, Universidade Federal Rural de Pernambuco,  52171-900, Recife --
PE, Brazil}
\author{L. C. de Souza}
\affiliation{Departamento de F\'isica, Universidade Federal Rural
de Pernambuco, 52171-900, Recife -- PE, Brazil}

\author{Andr\'e L. M. Vilela}
\affiliation{F\'isica de Materiais, Universidade de Pernambuco, 50720-001,
Recife -- PE, Brazil}
\affiliation{Center for Polymer Studies and Department of Physics, Boston University, 02215, Boston -- MA, USA}

\author{H. Eugene Stanley}%
\affiliation{Center for Polymer Studies and Department of Physics, Boston University, Boston, MA 02215, USA}

\author{A. J. F. de Souza}
\affiliation{Departamento de F\'i­sica, Universidade Federal Rural
de Pernambuco, 52171-900, Recife -- PE, Brazil}
\date{\today}

\begin{abstract}
We perform short-time Monte Carlo simulations to study the criticality of the isotropic two-state majority-vote model on cubic lattices of volume $N = L^3$, with $L$ up to $2048$. We obtain the precise location of the critical point by examining the scaling properties of a new auxiliary function $\Psi$. We perform finite-time scaling analysis to accurately calculate the whole set of critical exponents, including the dynamical critical exponent $z=2.027(9)$, and the initial slip exponent $\theta = 0.1081(1)$. Our results indicate that the majority-vote model in three dimensions belongs to the same universality class of the three-dimensional Ising model.
\end{abstract}


\maketitle

\section{\label{Intd} Introduction}

The short-time Monte Carlo simulations focus on the time series analysis concerning the transitory behavior of proper physical observables of the system \cite{ZhengRev1998, PhysRevE.99.052104}. In this work, we center our investigation for the short-time dynamics of the two-state majority-vote model for values of the parameter $q$ nearly its critical value $q_c$ \cite{Oliveira1992, Oliveira1993}. The noise parameter $q$ acts as a social temperature, driving the order-disorder phase transition of the system. In the vicinity of the critical point, the order parameter $M(t)$ follows the following scaling relation ansatz~\cite{Janssen1989,li1994MCST,ZhengRev1998}
\begin{equation}
\label{law01}
\langle M(t) \rangle = 
b^{-\beta / \nu} {\cal M}(b^{-z}t,b^{1/\nu}\varepsilon),
\end{equation}
where $\langle \cdots \rangle$ indicates average over different realizations of the dynamics,
$\varepsilon = (q/q_c-1)$, $b$ is a spatial scale factor, $\beta$ and $\nu$ are the critical exponents associated with the order parameter and with divergence of the correlation length, respectively.  Here $z$ denotes the dynamical critical exponent, while the order parameter $M(t)$ supports the scaling relation Eq.~(\ref{law01}) for a fully ordered initial microstate, i. e., $M(0) = 1$. Nevertheless, the scaling behavior of the order parameter is also observed for other initial
configurations. For the case of an initial microstate where the order parameter is nearly zero, another independent dynamic exponent~\cite{PhysRevB.40.304,Janssen1989}, the initial slip exponent $\theta$, delineates the dynamics of the system when $M(0) \simeq 0$.

From Eq.~(\ref{law01}), it follows that~\cite{AlbanoRev2011}
\begin{equation}
\label{magvst}
\langle M(t) \rangle = t^{-\beta / \nu z}
{\cal F}(t^{1/\nu z}\varepsilon),
\end{equation}
and precisely at the critical point, where $\varepsilon = 0$, the exponent $\beta / \nu z $ is estimated by simple regression analysis. Additionally, we use the same method to estimate $q_c$, considering that a straight line fits the data properly only when $q$ is close to its critical value $q_c$.
A strong feature of this method is to avoid issues induced by the critical slowing-down phenomena, usually present in Monte Carlo simulations. The evaluations of the short-time analysis are carried out in the early stages of the dynamics, thus evading the relaxation period for the system to reaches equilibrium.

Although its apparent simplicity, the above-sketched method
has been successfully applied to a variety of condensed matter
systems~\cite{Luo1998383,ISI:000268747900024,PhysRevB.89.144115,
PhysRevE.61.7204,PhysRevB.76.224407,PhysRevE.66.026130,
PhysRevE.72.036122,PhysRevB.89.144115,superlattice,
PhysRevE.60.3666}. For instance, the dimensional crossover for an Iron-Vanadium magnetic superlattice model observed when the inter- to intralayer exchange coupling ratio approaches zero~\cite{superlattice}. The function of the iron vacancy in the magnetic order of a $J_1-J_2$ Ising model~\cite{PhysRevE.87.022113} is a deeply non-trivial application of the short-time method. In a lattice-gauge theory application~\cite{Frigori20101388},
the static and dynamical critical exponents of
a $(2+1)$-dimensional gluodynamics of the $SU(2)$ gauge theory were obtained, where the universality hypothesis was verified.
Recently other applications and new developments of the short-time analysis method were also employed~\cite{AlbanoRev2011}.

In this work, we apply a novel technique that allows the systematic and accurate evaluation of the critical point $q_c$, obtained from the analysis of the Monte Carlo
simulations data \cite{PhysRevE.99.052104}. Furthermore, the method provides estimates for
all static and dynamical critical exponents. We illustrate our method in the study of the critical behavior of the majority-vote (MV) model with noise on cubic lattices~\cite{Oliveira1992,Oliveira1993}.

The critical behavior of the majority-vote model has been investigated by several techniques on two dimensions for distinct lattice structures~\cite{santos1995,PhysRevE.67.026104,PhysRevE.89.052109}.
We also remark a number of studies and generalizations of the 
majority-vote dynamics which shed light on our
comprehension of the non-equilibrium
statistical mechanics~\cite{tome1998,Vilela20126456, Vilela2020, Vilela2017, Fiore2018, crokidakis2016,crochikdakis2005,Stone2015437,Vilela20094171,
1742-5468-2008-04-P04021,0305-4470-32-2-003,0305-4470-24-20-015,
lsac,lima2014,felipe,*[{See also: }]  PhysRevE.88.032142}.

In the standard two-state majority-vote model~\cite{Oliveira1992,Oliveira1993}, the opinion dynamics follows the majority rule, and the opinion of an individual is represented by a spin variable $\sigma$, which assumes two values: $\pm 1$. A given spin adopts the state in which most of its interacting neighbors are with probability $1-q$, and the opposite state with probability $q$.
This system undergoes an order-disorder phase transition at a finite value of the noise parameter $q$, which belongs to the universality class 
of corresponding equilibrium Ising models in
two~\cite{Oliveira1992,Oliveira1993,santos1995, PhysRevE.57.108,YANG,RevModPhys.76.663}, and probably also in three~\cite{PhysRevE.86.041123} dimensions. These findings support
the conjecture~\cite{PhysRevLett.55.2527} that non-equilibrium
model systems with up-down symmetry and spin-flip dynamics are
in the same universality class of the equilibrium Ising model.

Motivated by the results of the recent studies~\cite{PhysRevE.86.041123}, obtained from long-time Monte Carlo simulations of moderately small lattices, we decide to investigate the MV model in three-dimensions. On account of finite-size effects, the authors consider corrections terms for scaling in their analysis.
On the other hand, previous results~\cite{YANG} suggested that the 
MV model in three dimensions could violate the conjecture we mention before.
We believe it is essential to support the findings of
reference~\cite{PhysRevE.86.041123}
with simulations on larger lattices, avoiding correction to the scaling, and with different strategies. We also provide the first estimates for both the dynamical critical and the initial slip exponents to this model in three dimensions.

We describe the short-time Monte Carlo data analysis in section \ref{theory}
along with the MV model description. In the same section, we point our
observations concerning the computational procedure and numerical techniques.
In section \ref{res} we present our results. We conclude
with a summary and final remarks in section \ref{conc}.

\section{\label{theory} Model, Theory and Simulation}

To each node of a fully periodic cubic lattice of side $L$, we associate a 
Ising-like spin variable $\sigma_i = \pm 1$. Such spin interacts with
its six nearest neighbors. As a result of this interaction,
the spin changes its state according
to the majority rule~\cite{Oliveira1992}.
During an elementary time step, a node $i$ is randomly selected and the spin $\sigma_i$ is
flipped with a probability given by
\begin{equation}
w(\sigma_i) = \frac{1}{2} \left[1 - (1-2q) \sigma_i {\cal S}
\left( \sum_{\delta = 1}^6 \sigma_{i+\delta} \right) \right],
\end{equation}
where ${\cal S}(x) = \mbox{sgn}(x)$ if $x \ne 0$ and ${\cal S}(0) = 0$.
The sum runs over the nearest neighbors of the spin $\sigma_i$.
We measure time in Monte Carlo steps (MCS),  which consists of 
$L \times L \times L$ such elementary moves.

In a short-time critical dynamic analysis,
one prepares the system in an initial state.
Then the system is released to evolve according to the 
prescribed dynamics for some value of the control parameter $q$
until a specific time.
The whole process is repeated a certain number of times in order
to obtain a smooth averaged time series.
To our purpose, it is convenient to start from a fully ordered
initial state and to follow the temporal behavior of the order
parameter. Therefore, at $t=0$, we set $\sigma_i = 1$ for all $i \in N$, where $N = L^3$.
Thus, the magnetization 
\begin{equation}
M(t) = \frac{1}{L^3} \sum_i \sigma_i,
\end{equation}
at $t=0$ is unity.

For the ordered phase, where $q < q_c$, the system evolves toward a steady state, and $\langle M(t,q) \rangle$
decays to a constant roughly independent of the system size.
On the contrary, for $q > q_c$ one expects $\langle M(t,q) \rangle$
to drop to a value of the order of $1/L$. The relaxation is exponential
for $q$ not too close to $q_c$~\cite{lsac}, and it turns into a power-law when
$q$ approaches $q_c$, c.f., Eq.~(\ref{magvst}).
The overall behavior shows up more clearly when plotted in a
double logarithmic scale, and it is enhanced through the following
auxiliary function~\cite{PhysRevE.99.052104}

\begin{equation}
\label{Psi}
\Psi(t,q) = \frac{\partial}{\partial \tau}
\ln{ \langle M(t,q) \rangle },
\end{equation}
where $\tau = \ln{(t)}$.
From Eq.~(\ref{magvst}), we have
\begin{equation}
\label{cumulant}
\Psi(t,q) =  -\frac{\beta}{\nu z} + t^{1/\nu z}\varepsilon
\widetilde{\Psi}(t^{1/\nu z}\varepsilon ),
\end{equation}
with $\widetilde{\Psi}(x)$ being an universal scaling function. 

Thus $\Psi(t,q)$ either goes to zero or tends
to $-\infty$ with the increasing of $t$ for $\varepsilon < 0$ or $\varepsilon > 0$, respectively.
On the other hand, at the critical point $\Psi(t,q=q_c) = -\beta /\nu z$, 
a constant independent of $t$ in the scaling regime. 
Furthermore, the famile of curves $\Psi(t_i,q)$, taken at distinct times $t_i$,
plotted against $q$ cross at the single point $(\beta/\nu z,q_c)$.
Having an estimate of the critical noise $q_c$, the exponent $1/\nu z$ can
be obtained by plotting the curves $\Psi(t_i,q)$ against the proper
scaling variable $x = t^{1/\nu z}\varepsilon$. Once all curves should
collapse onto a single curve only for the correct value of this
exponent when calculating $x$.
A similar scaling plot holds to the magnetization data. According to
Eq.~(\ref{magvst}), the plot of
$t^{\beta / \nu z} \langle M(t,q) \rangle $ against $x = t^{1/\nu z}\varepsilon$
also collapses onto a single curve for precise values of the critical parameters. Besides that, we can explore the
time evolution of the logarithmic derivative of the magnetization with respect to $\varepsilon$. The finite-time scaling law for the magnetization, Eq.~(\ref{magvst}), gives
\begin{equation}
\label{nu_z}
\partial_{\varepsilon} \ln{M(t,\varepsilon)}|_{\varepsilon = 0} \sim t^{1/\nu z}.
\end{equation}

Until now, we neglected any finite-size effects.
Right at the beginning of the time evolution
the fluctuations are spatially uncorrelated.
Thereby, the effective correlation length $\xi(t)$ is very small
for small $t$. As $t$ increases, $\xi(t)$ eventually becomes similar to
the equilibrium correlation length $\xi_{\text{eq}}(q)$.
From there on, the behavior of $\langle M(t,q) \rangle$ 
crosses over towards its steady-state value,
which is finite for finite $L$, even for $q$ above $q_c$.
In this way, $\Psi(t,q)$ goes to zero as $t \rightarrow \infty$ independently of
$q$. The inflection points in the $\Psi(t,q)$ curves are hallmarks of the crossover to equilibrium and a clear signal of finite-size effects.
The analysis must be carried out within a time window in which $\Psi(t,q)$ presents a monotonous behavior
with respect to $t$.

Besides the magnetization, we measure its second moment
\begin{equation}
M^{(2)} (t,q) = \frac{1}{L^3} \left\langle 
\left( \sum_i \sigma_i \right)^2 \right\rangle.
\end{equation}
That allows for calculating the fluctuation of the order parameter
\begin{equation}
\Delta M (t) =  \frac{1}{L^3} \left\langle 
\left( \sum_i \sigma_i \right)^2 \right\rangle
- \frac{1}{L^3} \left\langle 
\sum_i \sigma_i  \right\rangle^2,
\end{equation}
and the second-order cumulant~\cite{AlbanoRev2011,ZhengRev1998}

\begin{equation}
\label{eq_u2}
U_2(t) = \frac{ M^{(2)} }{ \langle M \rangle^2  } - 1.
\end{equation}

Let us now focus on the time evolution of
the magnetization starting from a disordered state.
The initial microstate has magnetization $m_0 << 1$
and negligible spatial correlation.
With this kind of initial condition,
the time evolution of the magnetization at the
critical noise becomes \cite{AlbanoRev2011}
$M(t) \propto t^{\theta}$, where the exponent $\theta$ 
controls the rate of growth of the magnetization for short times.
This power-law initial increase is observed only in
the limit $m_0 \rightarrow 0$. In practice, data must
be extrapolated for $m_0 = 0$.

In our numerical simulations, we use the parallel
computing platform CUDA~\cite{CUDA},
developed by NVIDIA corporation. Although we do not present the
implementation details, quite relevant information can be found in the reference by Preis and co-workers~\cite{pvps_jcp_2009}.
Here we present only a brief description of our procedure.

The CUDA programming model allows us to perform simulations in a massively parallel way. We obtain data on several cubic lattices in a single run on a Graphics Processing Unit (GPU). We increase the parallelism even further storing one spin in a single bit, i. e., $32$ spins per computer word~\cite{murilinho,PhysRevB.48.9586}.

To maximize parallelization, we stack several lattices on top of each other and update them simultaneously.
To avoid neighbor spins to be changed at the same time, 
we divide the lattice into $8$ sub-lattices, and perform a Monte 
Carlo step in $8$ iterations. In each iteration, all spins belonging to the same sub-lattice are updated in parallel. This procedure is not equivalent to randomly select one spin to flip at a time. Nevertheless, it affects only non-universal quantities, as the value of the phase transition point \cite{Odor2000}.

\section{\label{res} Critical relaxation of the three-dimensional
majority vote model}

We perform Monte Carlo simulations on three-dimensional lattices of linear sizes $L = 256, 512, 1024$, and $2048$,
with periodic boundary conditions. We focus our presentation for different initial conditions: (a) for $L = 256$ with a fully ordered initial state, and (b) for $L = 2048$ with a disordered initial configuration, where $m_0 \simeq 0$.

\subsection{Ordered initial state}

We investigate the time behavior for a variety of physical quantities
defined on a cubic lattice with periodic boundary conditions and side $L=256$. We choose a fully ordered initial state, and we record the magnetization up to $10^4$ MCS for several values of the noise $q$ near $q_c$. In each simulation, we generate $1024$ independent samples for a given noise value $q$. We reproduce the simulations using the same noise until we obtain data smooth enough to calculate reliable numerical derivatives. 

\begin{figure}
\includegraphics[width=0.5\textwidth]{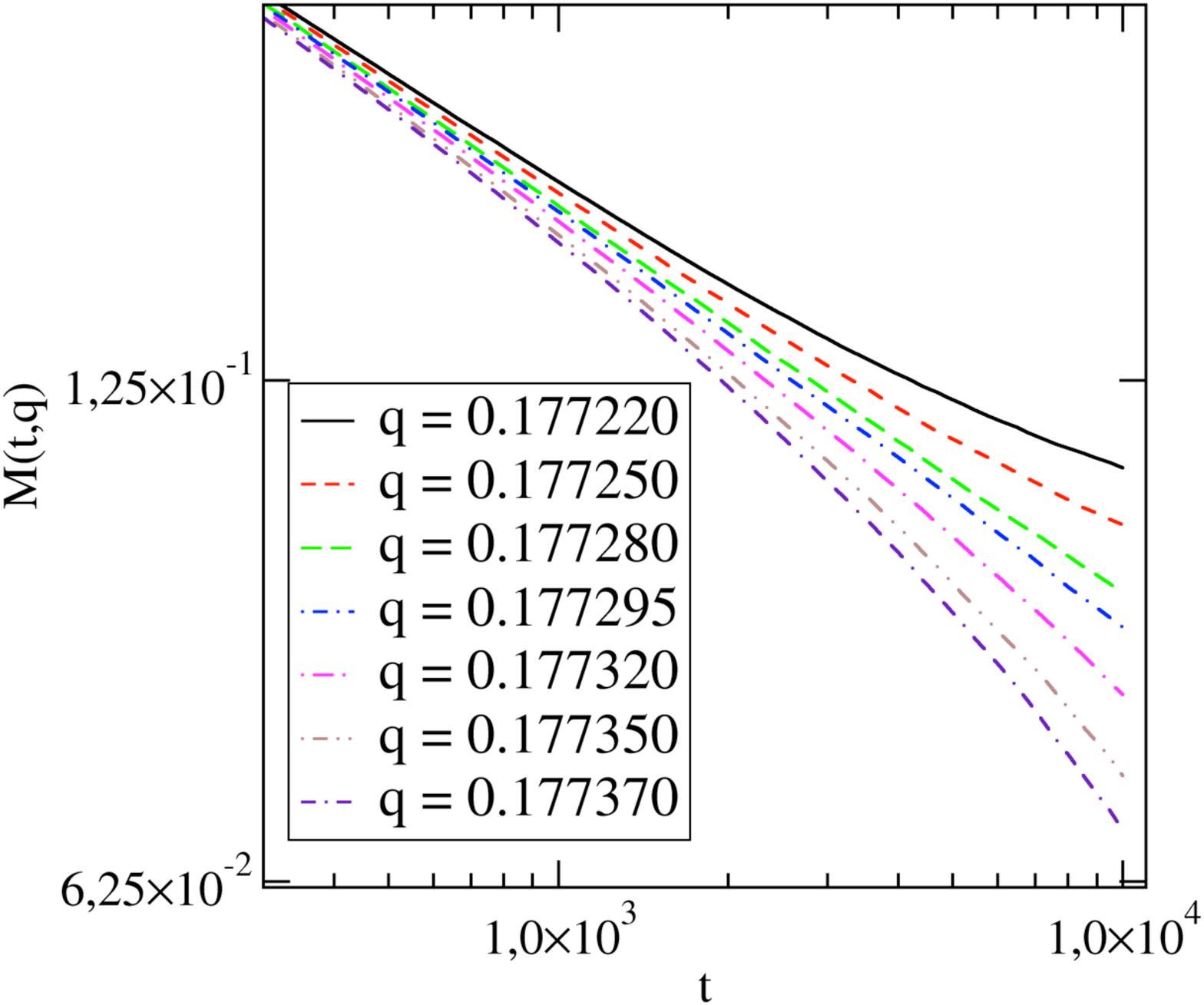}%
 \caption{\label{Mag} (Color online) Average magnetization as a function of time for $L = 256$ and several values of the noise near the critical
point $q_c = 0.177293$. Below the critical noise, it shows a trend to saturate in the long-time regime, indicating the presence of an ordered steady-state. The average magnetization displays a faster than power-law decay above the critical noise.}
\end{figure}

In Fig.~\ref{Mag}, we show the time evolution of the magnetization
averaged over at least $5120$ samples.
We report the representative results, although we carried out simulations for many noises in the range of $0.17722 \le q \le 0.17737$.
For the sake of clarity, we do not display the error bars.
For noises values below the critical point, the magnetization
relaxes from the initial non-equilibrium state
toward a finite amount. This behavior indicates the presence of long-range order. Above the critical noise, the curves lean
down, signaling that the system is in a disordered
phase. At the critical noise, the magnetization develops a
slow power-law decay after a microscopic transient time.

\begin{figure}
 \includegraphics[width=0.5\textwidth]{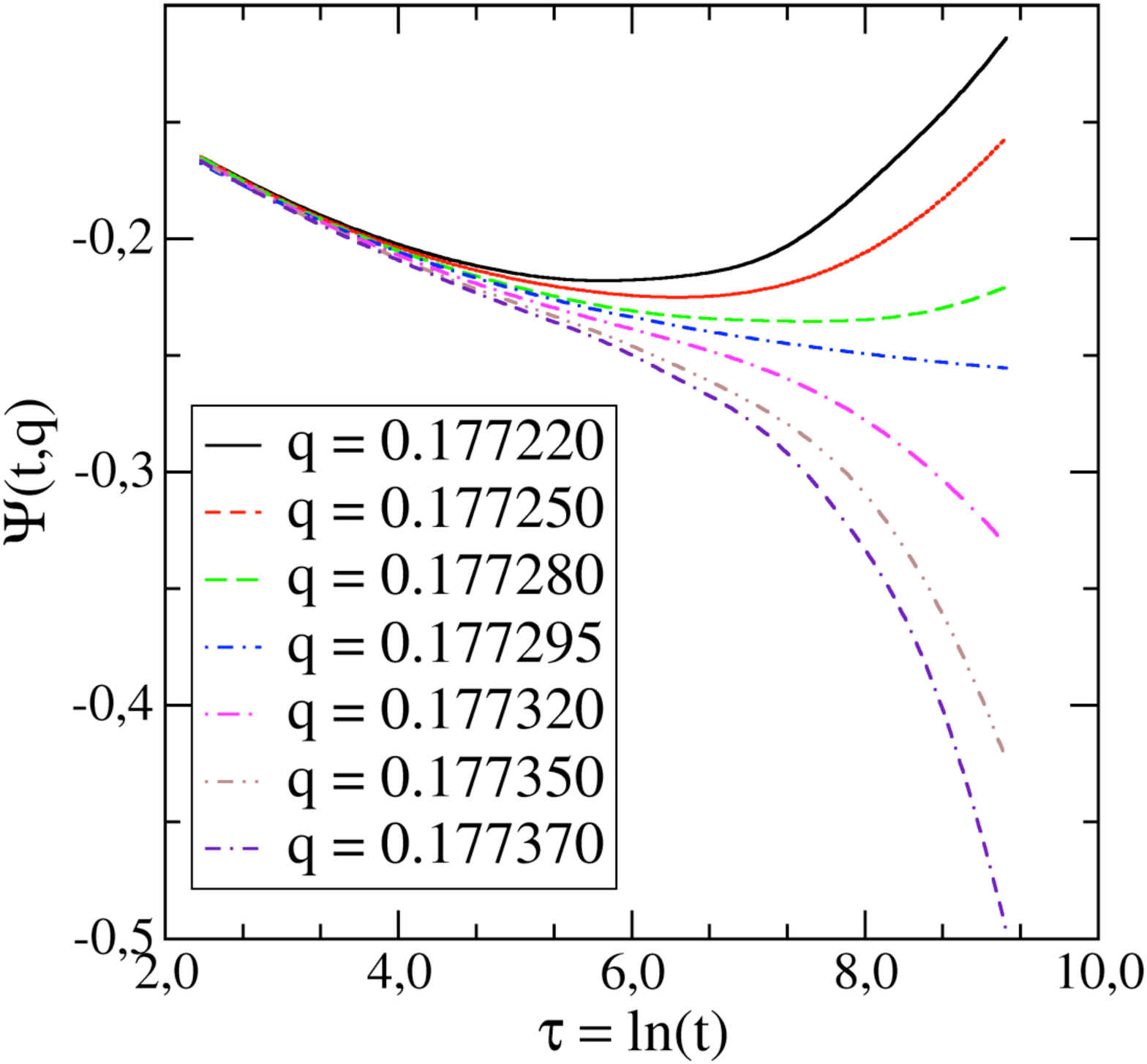}%
 \caption{\label{Psi1} (Color online) The auxiliary function
$\Psi(t,q)$ as a function of $\ln{(t)}$ in the vicinity of
the critical point $q_c = 0.177293$.}
\end{figure}

In Fig.~\ref{Psi1}, we show the auxiliary function $\Psi(t,q)$
as a function of $\ln{(t)}$ for the same noises as in Fig.~\ref{Mag}.
Now, one sees a clearer signature of the critical point.
For noises values below the critical one, the auxiliary function
goes to zero in the long-time regime due to the residual ordering
of the system. On the other hand, $\Psi(t,q)$ assumes diverging negative
values above the critical point, reflecting the exponential relaxation
towards the disordered steady-state. Precisely at the critical point, it
assumes a constant value after a microscopic transient time.
From the data of Fig.~\ref{Psi1}, it is possible infer that
the critical noise is close to $q = 0.17729$.

\begin{figure}
 \includegraphics[width=0.5\textwidth]{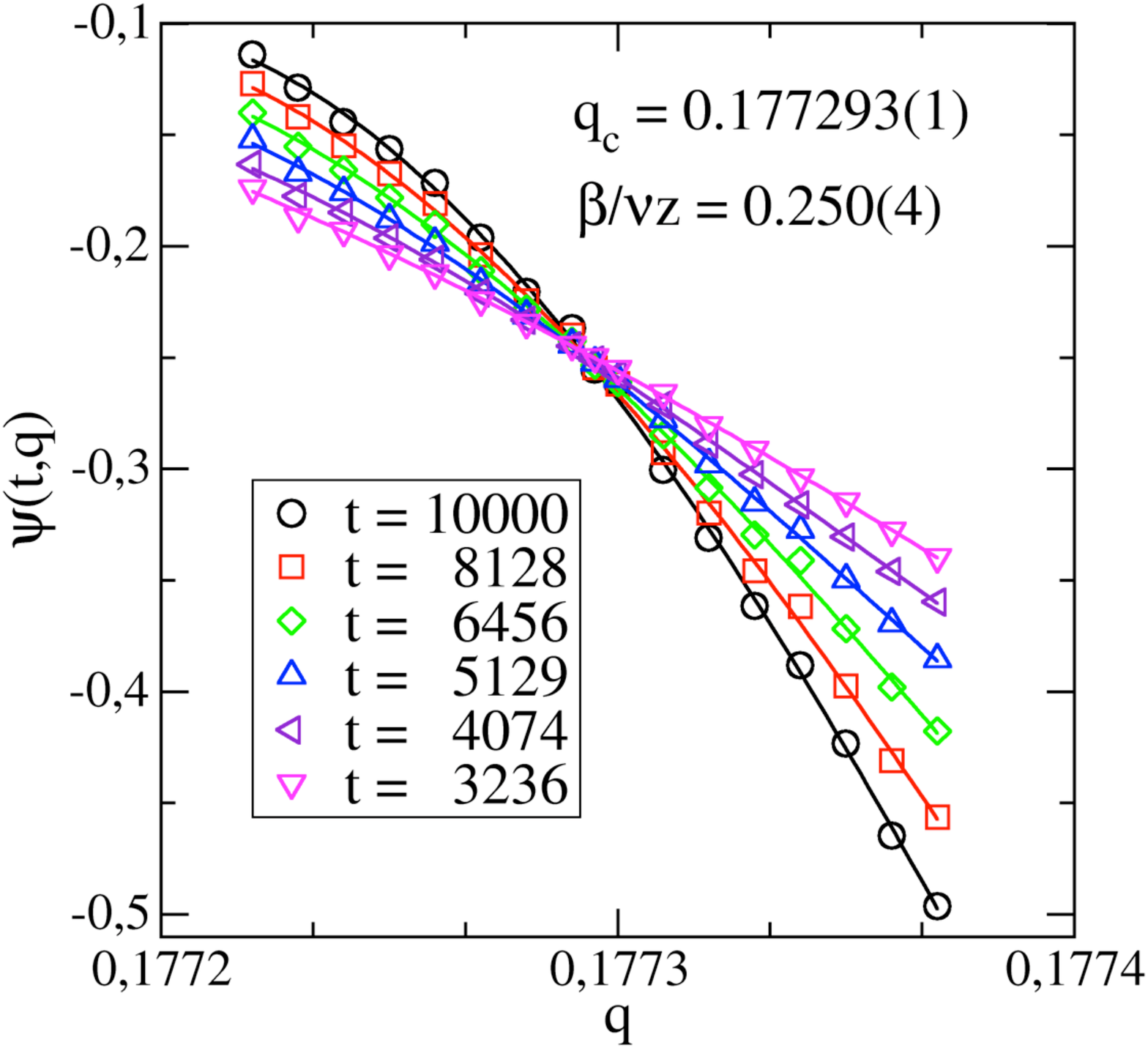}%
 \caption{\label{Psi2} (Color online) The auxiliary function 
$\Psi(t,q)$ versus $q$ taken at distinct times.
All curves intersect at a single point, identifying the
critical noise $q_c$ and the exponent ratio $\beta / \nu z$.
We estimate $q_c = 0.177293(1)$ and 
$\beta/ \nu z = 0.250(4)$.}
\end{figure}
As stated before, we can obtain a more precise estimate of the
critical point location by plotting the data of Fig.~\ref{Psi1}
in a different form, specifically, considering $\Psi(t,q)$
as a function of $q$ for a selected set of values
of $t$. In Fig.~\ref{Psi2}, we plot the auxiliary function
$\Psi(t,q)$, as a function of the noise $q$, for
a elected set of run times.
The curves have a common intersection point $(q_c,-\beta/ \nu z)$,
in which the auxiliary function does not depend on time $t$.
All curves cross at virtually one single point. The notably narrow
spread of the crossings is a definite indication that no relevant corrections to scaling are present in the data.
From these crossings, we estimate $q_c = 0.177293(1)$
and $\beta/ \nu z = 0.250(4)$.
We compare our estimate for $q_c$ with $q_c = 0.17628(7)$
from reference \cite{PhysRevE.86.041123}. The disagreement between the two estimates is due to the different updating schemes employed in each simulation.

\begin{figure}
\includegraphics[width=0.5\textwidth]{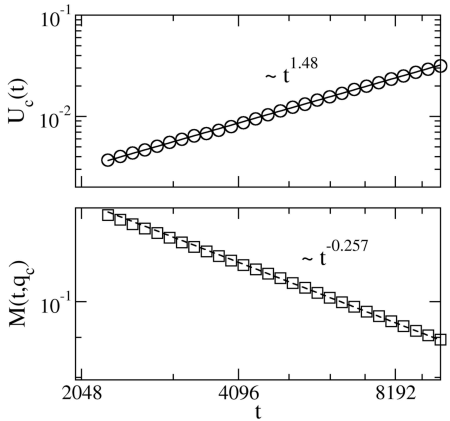}%
 \caption{\label{Fig_u2} (Color online)
(Top) Time-evolution of the second cumulant $U_2(t)$
calculated at the critical noise. Continuous line corresponds
to the power-law growth $U (t) \propto t^{3/z}$.
(Botton) Time-evolution of the magnetization at the
critical noise. The dashed line is consistent with the
power-law decaying $M(t) \propto t^{-\beta / \nu z}.$}
\end{figure}

Owning an accurate estimate of the critical noise, we can determine
the dynamic critical exponent $z$, from the temporal evolution of the
second-order cumulant defined by Eq.~(\ref{eq_u2}). To obtain a direct
estimative for $z$, we performe simulations at the critical noise
$q_c = 0.177293$ on lattices of side $L = 256$, for $10240$ independent samples.
According to the finite-time scaling behavior, the second
cumulant shall grow in time as $U_c (t) \propto t^{d/z}$ at the
critical point, where $d$ is the space dimension \cite{AlbanoRev2011}.
In the upper panel of Fig.~\ref{Fig_u2}, we present our data
for $U_2(t)$ at the critical noise $q_c$. From these data we got $3/z = 1.480(7)$.
Besides the exponent $z$, the new data provide an estimate
of $\beta / \nu z$ more accurate than that obtained
from the intersections of the curves in Fig.~\ref{Psi2},
owing to its superior statistical quality.
In the botton panel of Fig.~\ref{Fig_u2}, we report the time evolution of the magnetization at the critical point, from which
we accurately estimate $\beta / \nu z = 0.2567(9)$.

\begin{figure}
 \includegraphics[width=0.5\textwidth]{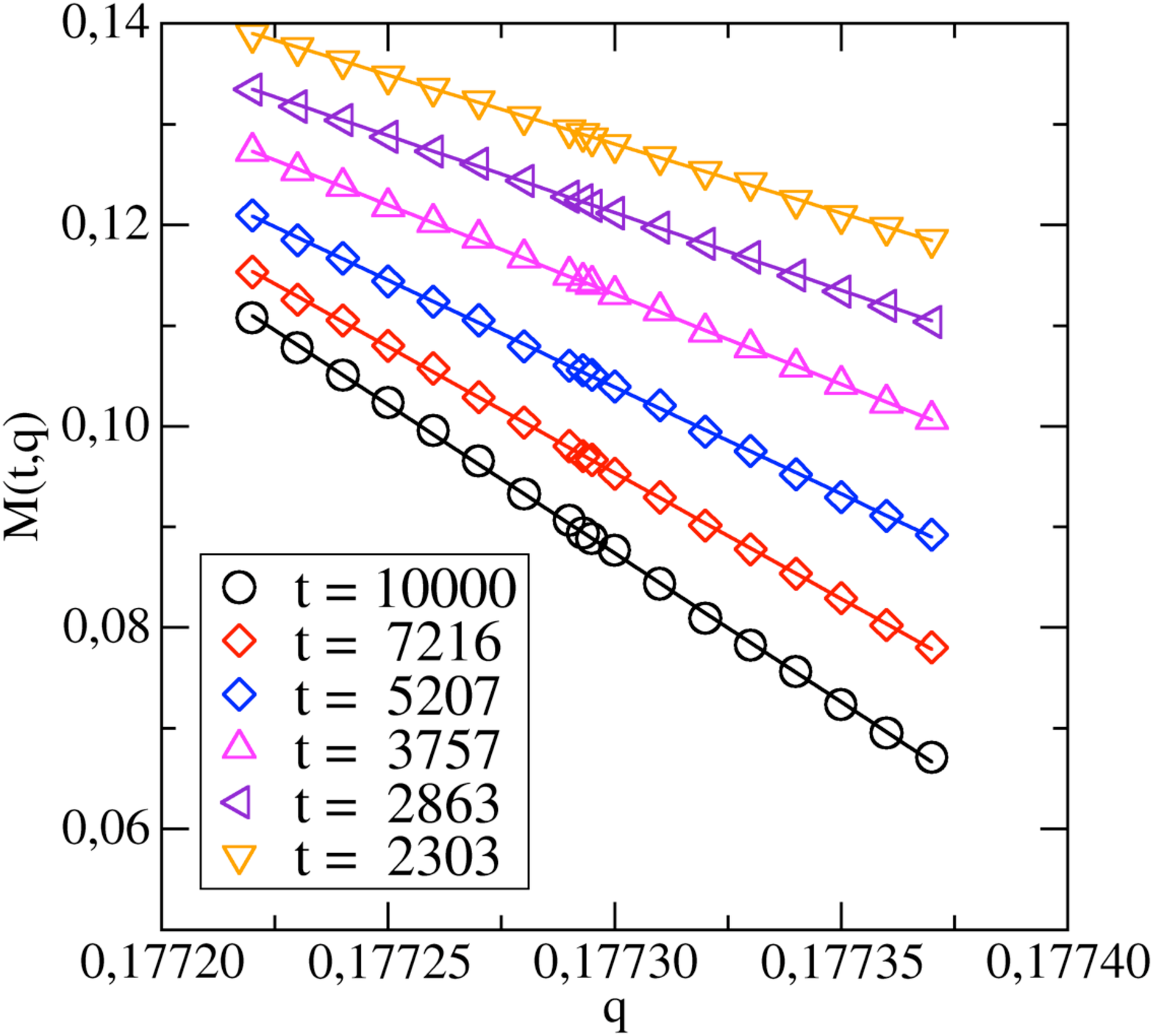}%
 \caption{\label{Mag5} (Color online)
Magnetization as a function of $q$ for several values of $t$.
The results are well fitted by straight lines. The error bars are smaller than the symbol sizes.} 
\end{figure}

To obtain an estimate of the $1/\nu z$, we notice from Fig.~\ref{Mag5} that the magnetization as a function of noise at a given time is fairly linear. Hence, the data corresponding to, say, $t = t_i$ is well described by
$M(t_i,q) = c_{0,i} + qc_{1,i}$, where $c_{0,i}$ and $c_{1,i}$
are calculated by a least-square method. Thus
\begin{equation}
\left. \partial_{q} \log{M(t,q)} \right |_{t=t_i} = 
\frac{c_{1,i}}{c_{0,i} + qc_{1,i}},
\end{equation}
for any $q$ in the range $[0.17722, 0.17737]$.

\begin{figure}
 \includegraphics[width=0.5\textwidth]{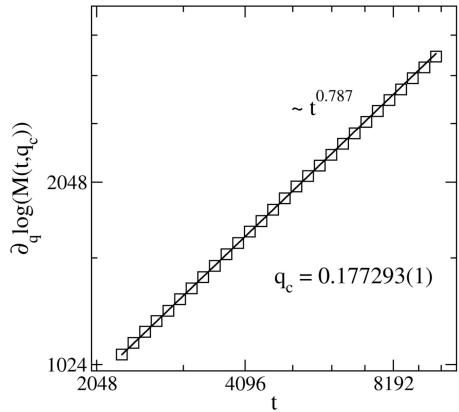}%
 \caption{\label{Mag6} (Color online)
Time evolution of the logarithmic derivative of the
magnetization at the critical noise $q_c =  0.177293(1)$.
Continuous line corresponds
to the power-law growth $\partial_q \log M(t) \propto t^{1/\nu z}$.}
\end{figure}

In Fig.~\ref{Mag6}, we exhibit the time evolution of the logarithmic derivative of the magnetization with respect to $q$ at the critical noise.
The slope of the straight line provides $1/ \nu z = 0.7869(16)$.

\begin{figure}
 \includegraphics[width=0.5\textwidth]{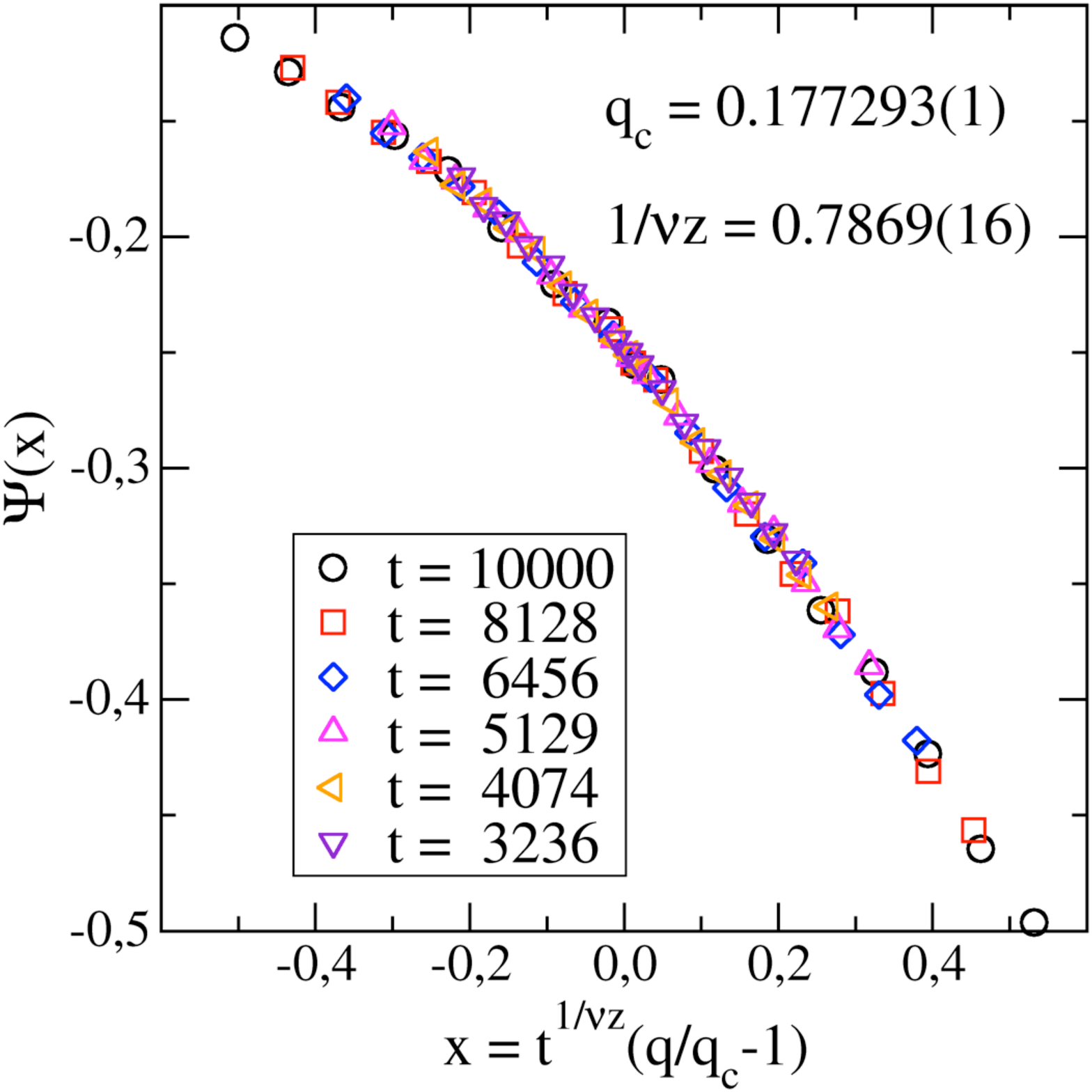}%
 \caption{\label{Psi_scaled} (Color online)
The auxiliary function data plotted against
the scaling variable $x = t^{\nu z}(q/q_c-1)$.}
\end{figure}

\begin{figure}
 \includegraphics[width=0.5\textwidth]{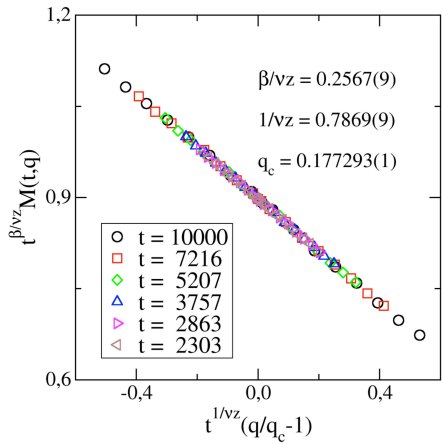}%
 \caption{\label{scaled_mag} (Color online)
Data collapse of the magnetization data computed
at distinct times and noises. The data collapse validates our estimates for the critical parameters $\beta / \nu z$, $1/\nu z$, and $q_c$.}
\end{figure}

According to the finite-time scaling hypothesis the
data from both Fig.~\ref{Psi2} and Fig.~\ref{Mag5} should collapse onto a single curve, provided that their
respective axes be properly rescaled.
This scaling analysis can be used to further verify
the precision of the above estimates for the critical
parameter of the three-dimensional MV model. 
In Fig.~\ref{Psi_scaled}, we plot our data for 
the auxiliary function $\Psi(t,q)$ as a function
of $x = t^{1/\nu z}(q/q_c-1)$. Similarly, we plot
$t^{\beta / \nu z}M(t,q)$ against $x$ in Fig~\ref{scaled_mag}.
The plots of Fig.~\ref{Psi_scaled} and Fig~\ref{scaled_mag}
show excellent agreement with the finite-time scaling assumption.
They also give evidence of the correctness
of our estimates for the critical parameters.

\subsection{Disordered initial state}

\begin{figure}
 \includegraphics[width=0.5\textwidth]{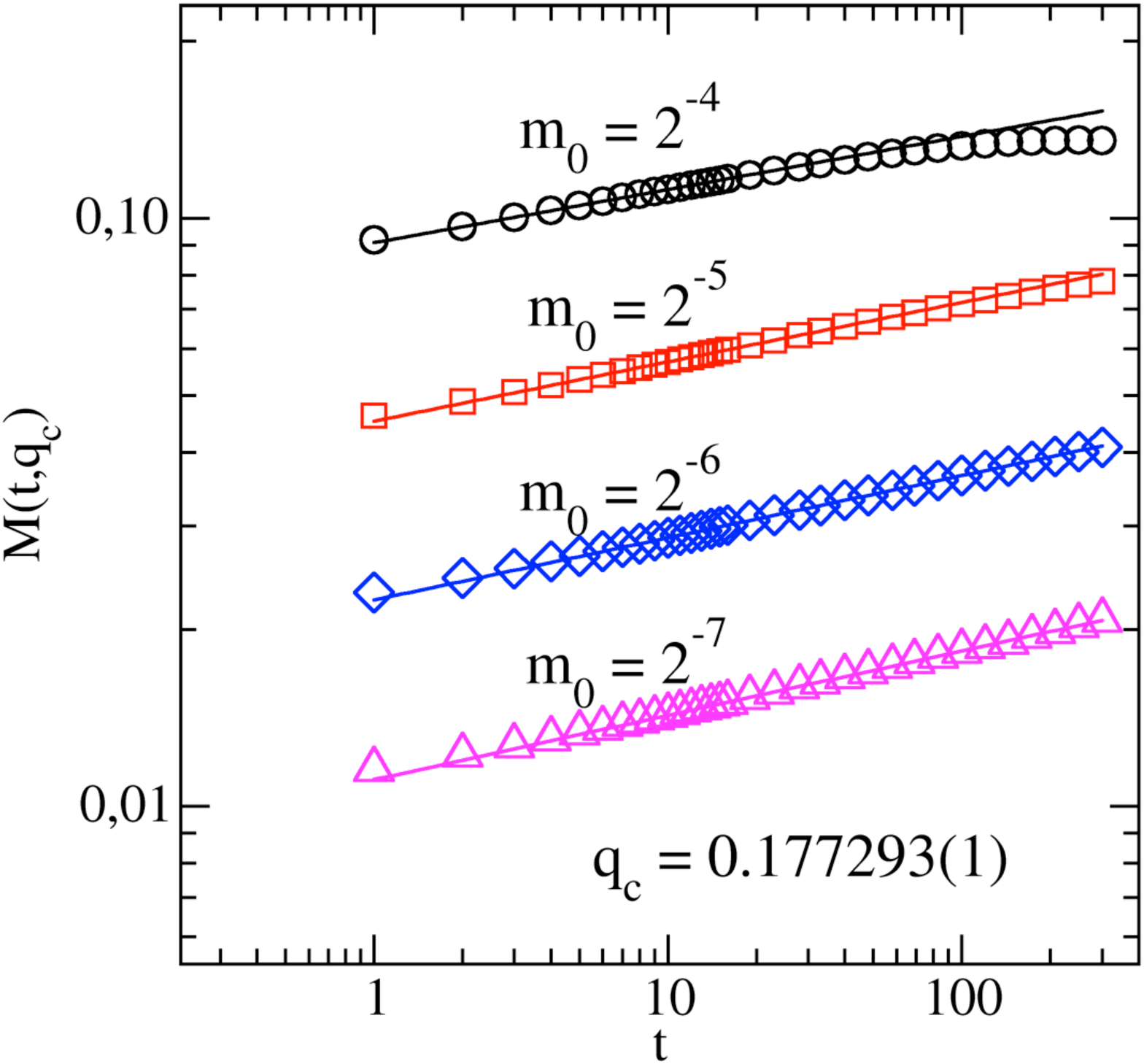}%
 \caption{\label{Mag4} (Color online)
Time evolution of the averaged 
magnetization at the critical noise $q_c =  0.177293(1)$
starting from disordered microstates.
Lines correspond to the power-law growth
 $ M(t) \propto t^{\theta}$. This result provides $\theta =  0.1081(1)$. }
\end{figure}

\begin{table*}
\caption{\label{m0_theta}
Critical parameters for the majority-vote and Ising models
in three dimensions.}
\begin{ruledtabular}
\begin{tabular}{dddddd}
m_0 & 0.00390625 & 0.0078125 & 0.015625 & 0.03125 & 0.0625 \\ 
\theta & 0.1082(3) &  0.1074(2)  & 0.10578(4) & 0.10086(9) & 0.0904(1)
\end{tabular}
\end{ruledtabular}
\end{table*}

To estimate the initial slip exponent, we simulate the short-time evolution of the magnetization in cubic lattices of side $L=2048$ for $300$ MCS, starting
from a disordered intial state. We measure the magnetization $M(t)$ for
$m_0 = 2^{-4}, 2^{-5},2^{-6},2^{-7}$, and $2^{-8}$.
For each case, we average $M(t)$ over $16$ initial
conditions and time history. We summarize our results in Fig.~\ref{Mag4}, where we display the time
evolution of magnetization for several values of $m_0$.
As shown in Table \ref{m0_theta}, the measured exponent $\theta$ depends on the initial magnetization $m_0$.
From these data, we apply the Bulirsch–Stoer (BST) extrapolation
method \cite{Alves_2000} to obtain $\theta =  0.1081(1)$ for $m_0 = 0$. We remark that models belonging to distinct stationary universality class can present the same dynamic initial slip exponent \cite{PhysRevE.96.052136}.

\section{\label{conc} Conclusions}

We have investigated the majority-vote model in three-dimensional cubic lattices using large-scale GPU Monte Carlo simulations. We accurately followed the short-time critical relaxation process from both fully ordered and disordered initial states. In our analysis, we use regular cubic lattices large enough for finite-size effects to be negligible during the simulation. Besides, we were able to investigate the deep scaling regime for a time interval that is sufficiently long to eliminate the need corrections to the scaling. We
obtain the critical parameters of the system by 
exploring the scaling properties of a new auxiliary  
function defined by Eq.~(\ref{Psi}) along with the 
order parameter. This function provides a precise location of the critical point of the system. Thus, we obtain the critical noise $q_c = 0.177293(1)$, associated with the critical exponent ratios $\beta / \nu z = 0.2567(9)$,
and $1/\nu z = 0.7869(16)$. In addition, we calculate the
dynamical critical exponent $z = 2.027(9)$ by
the time evolution of the second-order cumulant,
and the initial slip exponent $\theta = 0.1081(1)$ by
the initial increase of the magnetization (starting
from a disordered state).
From this set of exponents our results provide
the following estimates of the static critical
exponents $\beta = 0.3262(13)$ and $\nu = 0.627(4)$.
These values are in complete agreement with the three-dimensional
Ising universality class.
Recent large-scale Monte Carlo study of
a 3D Ising model yields $\nu = 0.629912(86)$\cite{PhysRevE.97.043301}.
Estimates based on field-theoretical methods provide
$\beta = 0.32645(6)$ and $\nu = 0.62999(5)$ \cite{LUNDOW201840}.
Our results also agree with the long-time Monte Carlo simulations of the MV model, where
$\beta = 0.331(34)$ and $\nu = 0.626(11)$ \cite{PhysRevE.86.041123}.
We believe this is the first work to obtain the dynamic critical exponents $\theta$ and $z$ for the majority-vote model in three-dimensional regular lattices.
Nevertheless, our findings are very close
to $z = 2.03(4)$ \cite{PhysRevB.43.6006}, 
$z = 2.0245(15)$ \cite{PhysRevE.101.022126}, and
$\theta = 0.108(2)$ \cite{Jaster}
from simulations of the three-dimensional Ising model.
Therefore, the majority-vote model in three-dimensions
belongs to the three-dimensional Ising universality class.

We remark that the method of analyzing data from short-time critical dynamics using the auxiliary function $\Psi$ is quite general, and adequately robust to investigate the critical behavior of further complex statistical systems.

\begin{acknowledgments}
To the bright memory of our wonderful and dedicated friend and teacher F. G. Brady Moreira, who recently passed away.

The authors acknowledge financial support from NVIDIA Data Science GPU Program, and the funding agencies FACEPE (APQ-0565-1.05/14, APQ-0707- 1.05/14), CAPES, and CNPq. The Boston University work was supported by NSF Grant PHY-1505000.
\end{acknowledgments}


\begin{thebibliography}{55}%
\makeatletter
\providecommand \@ifxundefined [1]{%
 \@ifx{#1\undefined}
}%
\providecommand \@ifnum [1]{%
 \ifnum #1\expandafter \@firstoftwo
 \else \expandafter \@secondoftwo
 \fi
}%
\providecommand \@ifx [1]{%
 \ifx #1\expandafter \@firstoftwo
 \else \expandafter \@secondoftwo
 \fi
}%
\providecommand \natexlab [1]{#1}%
\providecommand \enquote  [1]{``#1''}%
\providecommand \bibnamefont  [1]{#1}%
\providecommand \bibfnamefont [1]{#1}%
\providecommand \citenamefont [1]{#1}%
\providecommand \href@noop [0]{\@secondoftwo}%
\providecommand \href [0]{\begingroup \@sanitize@url \@href}%
\providecommand \@href[1]{\@@startlink{#1}\@@href}%
\providecommand \@@href[1]{\endgroup#1\@@endlink}%
\providecommand \@sanitize@url [0]{\catcode `\\12\catcode `\$12\catcode
  `\&12\catcode `\#12\catcode `\^12\catcode `\_12\catcode `\%12\relax}%
\providecommand \@@startlink[1]{}%
\providecommand \@@endlink[0]{}%
\providecommand \url  [0]{\begingroup\@sanitize@url \@url }%
\providecommand \@url [1]{\endgroup\@href {#1}{\urlprefix }}%
\providecommand \urlprefix  [0]{URL }%
\providecommand \Eprint [0]{\href }%
\providecommand \doibase [0]{http://dx.doi.org/}%
\providecommand \selectlanguage [0]{\@gobble}%
\providecommand \bibinfo  [0]{\@secondoftwo}%
\providecommand \bibfield  [0]{\@secondoftwo}%
\providecommand \translation [1]{[#1]}%
\providecommand \BibitemOpen [0]{}%
\providecommand \bibitemStop [0]{}%
\providecommand \bibitemNoStop [0]{.\EOS\space}%
\providecommand \EOS [0]{\spacefactor3000\relax}%
\providecommand \BibitemShut  [1]{\csname bibitem#1\endcsname}%
\let\auto@bib@innerbib\@empty
\bibitem [{\citenamefont {Zheng}(1998)}]{ZhengRev1998}%
  \BibitemOpen
  \bibfield  {author} {\bibinfo {author} {\bibfnamefont {B.}~\bibnamefont
  {Zheng}},\ }\href@noop {} {\bibfield  {journal} {\bibinfo  {journal}
  {International Journal of Modern Physics B}\ }\textbf {\bibinfo {volume}
  {12}},\ \bibinfo {pages} {1419} (\bibinfo {year} {1998})}\BibitemShut
  {NoStop}%
\bibitem [{\citenamefont {de~Souza}\ \emph {et~al.}(2019)\citenamefont
  {de~Souza}, \citenamefont {de~Souza},\ and\ \citenamefont
  {Lyra}}]{PhysRevE.99.052104}%
  \BibitemOpen
  \bibfield  {author} {\bibinfo {author} {\bibfnamefont {L.~C.}\ \bibnamefont
  {de~Souza}}, \bibinfo {author} {\bibfnamefont {A.~J.~F.}\ \bibnamefont
  {de~Souza}}, \ and\ \bibinfo {author} {\bibfnamefont {M.~L.}\ \bibnamefont
  {Lyra}},\ }\href {\doibase 10.1103/PhysRevE.99.052104} {\bibfield  {journal}
  {\bibinfo  {journal} {Phys. Rev. E}\ }\textbf {\bibinfo {volume} {99}},\
  \bibinfo {pages} {052104} (\bibinfo {year} {2019})}\BibitemShut {NoStop}%
\bibitem [{\citenamefont {de~Oliveira}(1992)}]{Oliveira1992}%
  \BibitemOpen
  \bibfield  {author} {\bibinfo {author} {\bibfnamefont {M.~J.}\ \bibnamefont
  {de~Oliveira}},\ }\href {\doibase 10.1007/BF01060069} {\bibfield  {journal}
  {\bibinfo  {journal} {Journal of Statistical Physics}\ }\textbf {\bibinfo
  {volume} {66}},\ \bibinfo {pages} {273} (\bibinfo {year} {1992})}\BibitemShut
  {NoStop}%
\bibitem [{\citenamefont {de~Oliveira}\ \emph {et~al.}(1993)\citenamefont
  {de~Oliveira}, \citenamefont {Mendes},\ and\ \citenamefont
  {Santos}}]{Oliveira1993}%
  \BibitemOpen
  \bibfield  {author} {\bibinfo {author} {\bibfnamefont {M.~J.}\ \bibnamefont
  {de~Oliveira}}, \bibinfo {author} {\bibfnamefont {J.~F.~F.}\ \bibnamefont
  {Mendes}}, \ and\ \bibinfo {author} {\bibfnamefont {M.~A.}\ \bibnamefont
  {Santos}},\ }\href@noop {} {\bibfield  {journal} {\bibinfo  {journal}
  {Journal of Physics A: Mathematical and General}\ }\textbf {\bibinfo {volume}
  {26}},\ \bibinfo {pages} {2317} (\bibinfo {year} {1993})}\BibitemShut
  {NoStop}%
\bibitem [{\citenamefont {Janssen}\ \emph {et~al.}(1989)\citenamefont
  {Janssen}, \citenamefont {Schaub},\ and\ \citenamefont
  {Schmittmann}}]{Janssen1989}%
  \BibitemOpen
  \bibfield  {author} {\bibinfo {author} {\bibfnamefont {H.~K.}\ \bibnamefont
  {Janssen}}, \bibinfo {author} {\bibfnamefont {B.}~\bibnamefont {Schaub}}, \
  and\ \bibinfo {author} {\bibfnamefont {B.}~\bibnamefont {Schmittmann}},\
  }\href@noop {} {\bibfield  {journal} {\bibinfo  {journal} {Zeitschrift
  f{\"u}r Physik B Condensed Matter}\ }\textbf {\bibinfo {volume} {73}},\
  \bibinfo {pages} {539} (\bibinfo {year} {1989})}\BibitemShut {NoStop}%
\bibitem [{\citenamefont {Li}\ \emph {et~al.}(1994)\citenamefont {Li},
  \citenamefont {Ritschel},\ and\ \citenamefont {Zheng}}]{li1994MCST}%
  \BibitemOpen
  \bibfield  {author} {\bibinfo {author} {\bibfnamefont {Z.}~\bibnamefont
  {Li}}, \bibinfo {author} {\bibfnamefont {U.}~\bibnamefont {Ritschel}}, \ and\
  \bibinfo {author} {\bibfnamefont {B.}~\bibnamefont {Zheng}},\ }\href@noop {}
  {\bibfield  {journal} {\bibinfo  {journal} {Journal of Physics A:
  Mathematical and General}\ }\textbf {\bibinfo {volume} {27}},\ \bibinfo
  {pages} {L837} (\bibinfo {year} {1994})}\BibitemShut {NoStop}%
\bibitem [{\citenamefont {Huse}(1989)}]{PhysRevB.40.304}%
  \BibitemOpen
  \bibfield  {author} {\bibinfo {author} {\bibfnamefont {D.~A.}\ \bibnamefont
  {Huse}},\ }\href {\doibase 10.1103/PhysRevB.40.304} {\bibfield  {journal}
  {\bibinfo  {journal} {Phys. Rev. B}\ }\textbf {\bibinfo {volume} {40}},\
  \bibinfo {pages} {304} (\bibinfo {year} {1989})}\BibitemShut {NoStop}%
\bibitem [{\citenamefont {Albano}\ \emph {et~al.}(2011)\citenamefont {Albano},
  \citenamefont {Bab}, \citenamefont {Baglietto}, \citenamefont {Borzi},
  \citenamefont {Grigera}, \citenamefont {Loscar}, \citenamefont {Rodriguez},
  \citenamefont {Puzzo},\ and\ \citenamefont {Saracco}}]{AlbanoRev2011}%
  \BibitemOpen
  \bibfield  {author} {\bibinfo {author} {\bibfnamefont {E.~V.}\ \bibnamefont
  {Albano}}, \bibinfo {author} {\bibfnamefont {M.~A.}\ \bibnamefont {Bab}},
  \bibinfo {author} {\bibfnamefont {G.}~\bibnamefont {Baglietto}}, \bibinfo
  {author} {\bibfnamefont {R.~A.}\ \bibnamefont {Borzi}}, \bibinfo {author}
  {\bibfnamefont {T.~S.}\ \bibnamefont {Grigera}}, \bibinfo {author}
  {\bibfnamefont {E.~S.}\ \bibnamefont {Loscar}}, \bibinfo {author}
  {\bibfnamefont {D.~E.}\ \bibnamefont {Rodriguez}}, \bibinfo {author}
  {\bibfnamefont {M.~L.~R.}\ \bibnamefont {Puzzo}}, \ and\ \bibinfo {author}
  {\bibfnamefont {G.~P.}\ \bibnamefont {Saracco}},\ }\href@noop {} {\bibfield
  {journal} {\bibinfo  {journal} {Reports on Progress in Physics}\ }\textbf
  {\bibinfo {volume} {74}},\ \bibinfo {pages} {026501} (\bibinfo {year}
  {2011})}\BibitemShut {NoStop}%
\bibitem [{\citenamefont {Luo}\ \emph {et~al.}(1998)\citenamefont {Luo},
  \citenamefont {Schulz}, \citenamefont {Sch{\"u}lke}, \citenamefont
  {Trimper},\ and\ \citenamefont {Zheng}}]{Luo1998383}%
  \BibitemOpen
  \bibfield  {author} {\bibinfo {author} {\bibfnamefont {H.~J.}\ \bibnamefont
  {Luo}}, \bibinfo {author} {\bibfnamefont {M.}~\bibnamefont {Schulz}},
  \bibinfo {author} {\bibfnamefont {L.}~\bibnamefont {Sch{\"u}lke}}, \bibinfo
  {author} {\bibfnamefont {S.}~\bibnamefont {Trimper}}, \ and\ \bibinfo
  {author} {\bibfnamefont {B.}~\bibnamefont {Zheng}},\ }\href {\doibase
  http://dx.doi.org/10.1016/S0375-9601(98)80003-0} {\bibfield  {journal}
  {\bibinfo  {journal} {Physics Letters A}\ }\textbf {\bibinfo {volume}
  {250}},\ \bibinfo {pages} {383 } (\bibinfo {year} {1998})}\BibitemShut
  {NoStop}%
\bibitem [{\citenamefont {da~Silva}\ \emph {et~al.}(2009)\citenamefont
  {da~Silva}, \citenamefont {Fulco},\ and\ \citenamefont
  {Nobre}}]{ISI:000268747900024}%
  \BibitemOpen
  \bibfield  {author} {\bibinfo {author} {\bibfnamefont {L.~F.}\ \bibnamefont
  {da~Silva}}, \bibinfo {author} {\bibfnamefont {U.~L.}\ \bibnamefont {Fulco}},
  \ and\ \bibinfo {author} {\bibfnamefont {F.~D.}\ \bibnamefont {Nobre}},\
  }\href@noop {} {\bibfield  {journal} {\bibinfo  {journal} {{J. Phys.:
  Condens. Matter}}\ }\textbf {\bibinfo {volume} {{21}}},\ \bibinfo {pages}
  {{346005}} (\bibinfo {year} {{2009}})}\BibitemShut {NoStop}%
\bibitem [{\citenamefont {Yin}\ \emph {et~al.}(2014)\citenamefont {Yin},
  \citenamefont {Mai},\ and\ \citenamefont {Zhong}}]{PhysRevB.89.144115}%
  \BibitemOpen
  \bibfield  {author} {\bibinfo {author} {\bibfnamefont {S.}~\bibnamefont
  {Yin}}, \bibinfo {author} {\bibfnamefont {P.}~\bibnamefont {Mai}}, \ and\
  \bibinfo {author} {\bibfnamefont {F.}~\bibnamefont {Zhong}},\ }\href
  {\doibase 10.1103/PhysRevB.89.144115} {\bibfield  {journal} {\bibinfo
  {journal} {Phys. Rev. B}\ }\textbf {\bibinfo {volume} {89}},\ \bibinfo
  {pages} {144115} (\bibinfo {year} {2014})}\BibitemShut {NoStop}%
\bibitem [{\citenamefont {Santos}(2000)}]{PhysRevE.61.7204}%
  \BibitemOpen
  \bibfield  {author} {\bibinfo {author} {\bibfnamefont {M.}~\bibnamefont
  {Santos}},\ }\href {\doibase 10.1103/PhysRevE.61.7204} {\bibfield  {journal}
  {\bibinfo  {journal} {Phys. Rev. E}\ }\textbf {\bibinfo {volume} {61}},\
  \bibinfo {pages} {7204} (\bibinfo {year} {2000})}\BibitemShut {NoStop}%
\bibitem [{\citenamefont {Zelli}\ \emph {et~al.}(2007)\citenamefont {Zelli},
  \citenamefont {Boese},\ and\ \citenamefont {Southern}}]{PhysRevB.76.224407}%
  \BibitemOpen
  \bibfield  {author} {\bibinfo {author} {\bibfnamefont {M.}~\bibnamefont
  {Zelli}}, \bibinfo {author} {\bibfnamefont {K.}~\bibnamefont {Boese}}, \ and\
  \bibinfo {author} {\bibfnamefont {B.~W.}\ \bibnamefont {Southern}},\ }\href
  {\doibase 10.1103/PhysRevB.76.224407} {\bibfield  {journal} {\bibinfo
  {journal} {Phys. Rev. B}\ }\textbf {\bibinfo {volume} {76}},\ \bibinfo
  {pages} {224407} (\bibinfo {year} {2007})}\BibitemShut {NoStop}%
\bibitem [{\citenamefont {da~Silva}\ \emph {et~al.}(2002)\citenamefont
  {da~Silva}, \citenamefont {Alves},\ and\ \citenamefont {Drugowich~de
  Felicio}}]{PhysRevE.66.026130}%
  \BibitemOpen
  \bibfield  {author} {\bibinfo {author} {\bibfnamefont {R.}~\bibnamefont
  {da~Silva}}, \bibinfo {author} {\bibfnamefont {N.~A.}\ \bibnamefont {Alves}},
  \ and\ \bibinfo {author} {\bibfnamefont {J.~R.}\ \bibnamefont {Drugowich~de
  Felicio}},\ }\href {\doibase 10.1103/PhysRevE.66.026130} {\bibfield
  {journal} {\bibinfo  {journal} {Phys. Rev. E}\ }\textbf {\bibinfo {volume}
  {66}},\ \bibinfo {pages} {026130} (\bibinfo {year} {2002})}\BibitemShut
  {NoStop}%
\bibitem [{\citenamefont {Yin}\ \emph {et~al.}(2005)\citenamefont {Yin},
  \citenamefont {Zheng},\ and\ \citenamefont {Trimper}}]{PhysRevE.72.036122}%
  \BibitemOpen
  \bibfield  {author} {\bibinfo {author} {\bibfnamefont {J.~Q.}\ \bibnamefont
  {Yin}}, \bibinfo {author} {\bibfnamefont {B.}~\bibnamefont {Zheng}}, \ and\
  \bibinfo {author} {\bibfnamefont {S.}~\bibnamefont {Trimper}},\ }\href
  {\doibase 10.1103/PhysRevE.72.036122} {\bibfield  {journal} {\bibinfo
  {journal} {Phys. Rev. E}\ }\textbf {\bibinfo {volume} {72}},\ \bibinfo
  {pages} {036122} (\bibinfo {year} {2005})}\BibitemShut {NoStop}%
\bibitem [{\citenamefont {Murtazaev}\ and\ \citenamefont
  {Mutailamov}(2013)}]{superlattice}%
  \BibitemOpen
  \bibfield  {author} {\bibinfo {author} {\bibfnamefont {A.~K.}\ \bibnamefont
  {Murtazaev}}\ and\ \bibinfo {author} {\bibfnamefont {V.~A.}\ \bibnamefont
  {Mutailamov}},\ }\href {\doibase 10.1134/S1063776113040079} {\bibfield
  {journal} {\bibinfo  {journal} {Journal of Experimental and Theoretical
  Physics}\ }\textbf {\bibinfo {volume} {116}},\ \bibinfo {pages} {604}
  (\bibinfo {year} {2013})}\BibitemShut {NoStop}%
\bibitem [{\citenamefont {Brunstein}\ and\ \citenamefont
  {Tom\'e}(1999)}]{PhysRevE.60.3666}%
  \BibitemOpen
  \bibfield  {author} {\bibinfo {author} {\bibfnamefont {A.}~\bibnamefont
  {Brunstein}}\ and\ \bibinfo {author} {\bibfnamefont {T.}~\bibnamefont
  {Tom\'e}},\ }\href {\doibase 10.1103/PhysRevE.60.3666} {\bibfield  {journal}
  {\bibinfo  {journal} {Phys. Rev. E}\ }\textbf {\bibinfo {volume} {60}},\
  \bibinfo {pages} {3666} (\bibinfo {year} {1999})}\BibitemShut {NoStop}%
\bibitem [{\citenamefont {Zhou}\ \emph {et~al.}(2013)\citenamefont {Zhou},
  \citenamefont {Zheng},\ and\ \citenamefont {Dai}}]{PhysRevE.87.022113}%
  \BibitemOpen
  \bibfield  {author} {\bibinfo {author} {\bibfnamefont {N.~J.}\ \bibnamefont
  {Zhou}}, \bibinfo {author} {\bibfnamefont {B.}~\bibnamefont {Zheng}}, \ and\
  \bibinfo {author} {\bibfnamefont {J.~H.}\ \bibnamefont {Dai}},\ }\href
  {\doibase 10.1103/PhysRevE.87.022113} {\bibfield  {journal} {\bibinfo
  {journal} {Phys. Rev. E}\ }\textbf {\bibinfo {volume} {87}},\ \bibinfo
  {pages} {022113} (\bibinfo {year} {2013})}\BibitemShut {NoStop}%
\bibitem [{\citenamefont {Frigori}(2010)}]{Frigori20101388}%
  \BibitemOpen
  \bibfield  {author} {\bibinfo {author} {\bibfnamefont {R.~B.}\ \bibnamefont
  {Frigori}},\ }\href {\doibase http://dx.doi.org/10.1016/j.cpc.2010.04.005}
  {\bibfield  {journal} {\bibinfo  {journal} {Computer Physics Communications}\
  }\textbf {\bibinfo {volume} {181}},\ \bibinfo {pages} {1388 } (\bibinfo
  {year} {2010})}\BibitemShut {NoStop}%
\bibitem [{\citenamefont {Santos}\ and\ \citenamefont
  {Teixeira}(1995)}]{santos1995}%
  \BibitemOpen
  \bibfield  {author} {\bibinfo {author} {\bibfnamefont {M.~A.}\ \bibnamefont
  {Santos}}\ and\ \bibinfo {author} {\bibfnamefont {S.}~\bibnamefont
  {Teixeira}},\ }\href@noop {} {\bibfield  {journal} {\bibinfo  {journal}
  {Journal of statistical physics}\ }\textbf {\bibinfo {volume} {78}},\
  \bibinfo {pages} {963} (\bibinfo {year} {1995})}\BibitemShut {NoStop}%
\bibitem [{\citenamefont {Campos}\ \emph {et~al.}(2003)\citenamefont {Campos},
  \citenamefont {de~Oliveira},\ and\ \citenamefont
  {Moreira}}]{PhysRevE.67.026104}%
  \BibitemOpen
  \bibfield  {author} {\bibinfo {author} {\bibfnamefont {P.~R.~A.}\
  \bibnamefont {Campos}}, \bibinfo {author} {\bibfnamefont {V.~M.}\
  \bibnamefont {de~Oliveira}}, \ and\ \bibinfo {author} {\bibfnamefont
  {F.~G.~B.}\ \bibnamefont {Moreira}},\ }\href {\doibase
  10.1103/PhysRevE.67.026104} {\bibfield  {journal} {\bibinfo  {journal} {Phys.
  Rev. E}\ }\textbf {\bibinfo {volume} {67}},\ \bibinfo {pages} {026104}
  (\bibinfo {year} {2003})}\BibitemShut {NoStop}%
\bibitem [{\citenamefont {{Acu\~na-Lara}}\ \emph {et~al.}(2014)\citenamefont
  {{Acu\~na-Lara}}, \citenamefont {Sastre},\ and\ \citenamefont
  {Vargas-Arriola}}]{PhysRevE.89.052109}%
  \BibitemOpen
  \bibfield  {author} {\bibinfo {author} {\bibfnamefont {A.~L.}\ \bibnamefont
  {{Acu\~na-Lara}}}, \bibinfo {author} {\bibfnamefont {F.}~\bibnamefont
  {Sastre}}, \ and\ \bibinfo {author} {\bibfnamefont {J.~R.}\ \bibnamefont
  {Vargas-Arriola}},\ }\href {\doibase 10.1103/PhysRevE.89.052109} {\bibfield
  {journal} {\bibinfo  {journal} {Phys. Rev. E}\ }\textbf {\bibinfo {volume}
  {89}},\ \bibinfo {pages} {052109} (\bibinfo {year} {2014})}\BibitemShut
  {NoStop}%
\bibitem [{\citenamefont {Tom{\'e}}\ and\ \citenamefont
  {de~Oliveira}(1998)}]{tome1998}%
  \BibitemOpen
  \bibfield  {author} {\bibinfo {author} {\bibfnamefont {T.}~\bibnamefont
  {Tom{\'e}}}\ and\ \bibinfo {author} {\bibfnamefont {M.~J.}\ \bibnamefont
  {de~Oliveira}},\ }\href@noop {} {\bibfield  {journal} {\bibinfo  {journal}
  {Physical Review E}\ }\textbf {\bibinfo {volume} {58}},\ \bibinfo {pages}
  {4242} (\bibinfo {year} {1998})}\BibitemShut {NoStop}%
\bibitem [{\citenamefont {Vilela}\ \emph {et~al.}(2012)\citenamefont {Vilela},
  \citenamefont {Moreira},\ and\ \citenamefont {de~Souza}}]{Vilela20126456}%
  \BibitemOpen
  \bibfield  {author} {\bibinfo {author} {\bibfnamefont {A.~L.~M.}\
  \bibnamefont {Vilela}}, \bibinfo {author} {\bibfnamefont {F.~G.~B.}\
  \bibnamefont {Moreira}}, \ and\ \bibinfo {author} {\bibfnamefont {A.~J.~F.}\
  \bibnamefont {de~Souza}},\ }\href {\doibase
  http://dx.doi.org/10.1016/j.physa.2012.07.068} {\bibfield  {journal}
  {\bibinfo  {journal} {Physica A: Statistical Mechanics and its Applications}\
  }\textbf {\bibinfo {volume} {391}},\ \bibinfo {pages} {6456 } (\bibinfo
  {year} {2012})}\BibitemShut {NoStop}%
\bibitem [{\citenamefont {Vilela}\ \emph {et~al.}(2020)\citenamefont {Vilela},
  \citenamefont {Zubillaga}, \citenamefont {Wang}, \citenamefont {Wang},
  \citenamefont {Du},\ and\ \citenamefont {Stanley}}]{Vilela2020}%
  \BibitemOpen
  \bibfield  {author} {\bibinfo {author} {\bibfnamefont {A.~L.~M.}\
  \bibnamefont {Vilela}}, \bibinfo {author} {\bibfnamefont {B.~J.}\
  \bibnamefont {Zubillaga}}, \bibinfo {author} {\bibfnamefont {C.}~\bibnamefont
  {Wang}}, \bibinfo {author} {\bibfnamefont {M.}~\bibnamefont {Wang}}, \bibinfo
  {author} {\bibfnamefont {R.}~\bibnamefont {Du}}, \ and\ \bibinfo {author}
  {\bibfnamefont {H.~E.}\ \bibnamefont {Stanley}},\ }\href@noop {} {\bibfield
  {journal} {\bibinfo  {journal} {Scientific Reports}\ }\textbf {\bibinfo
  {volume} {10}},\ \bibinfo {pages} {8255} (\bibinfo {year}
  {2020})}\BibitemShut {NoStop}%
\bibitem [{\citenamefont {Vilela}\ and\ \citenamefont
  {de~Souza}(2017)}]{Vilela2017}%
  \BibitemOpen
  \bibfield  {author} {\bibinfo {author} {\bibfnamefont {A.~L.~M.}\
  \bibnamefont {Vilela}}\ and\ \bibinfo {author} {\bibfnamefont {A.~J.~F.}\
  \bibnamefont {de~Souza}},\ }\href@noop {} {\bibfield  {journal} {\bibinfo
  {journal} {Physica A}\ }\textbf {\bibinfo {volume} {488}},\ \bibinfo {pages}
  {216} (\bibinfo {year} {2017})}\BibitemShut {NoStop}%
\bibitem [{\citenamefont {de~Oliveira}\ \emph {et~al.}(2018)\citenamefont
  {de~Oliveira}, \citenamefont {da~Luz},\ and\ \citenamefont
  {Fiore}}]{Fiore2018}%
  \BibitemOpen
  \bibfield  {author} {\bibinfo {author} {\bibfnamefont {M.~M.}\ \bibnamefont
  {de~Oliveira}}, \bibinfo {author} {\bibfnamefont {M.~G.~E.}\ \bibnamefont
  {da~Luz}}, \ and\ \bibinfo {author} {\bibfnamefont {C.~E.}\ \bibnamefont
  {Fiore}},\ }\href@noop {} {\bibfield  {journal} {\bibinfo  {journal} {Physics
  Review E}\ }\textbf {\bibinfo {volume} {97}},\ \bibinfo {pages} {060101}
  (\bibinfo {year} {2018})}\BibitemShut {NoStop}%
\bibitem [{\citenamefont {Vieira}\ and\ \citenamefont
  {Crokidakis}(2016)}]{crokidakis2016}%
  \BibitemOpen
  \bibfield  {author} {\bibinfo {author} {\bibfnamefont {A.~R.}\ \bibnamefont
  {Vieira}}\ and\ \bibinfo {author} {\bibfnamefont {N.}~\bibnamefont
  {Crokidakis}},\ }\href@noop {} {\bibfield  {journal} {\bibinfo  {journal}
  {Physica A: Statistical Mechanics and its Applications}\ }\textbf {\bibinfo
  {volume} {450}},\ \bibinfo {pages} {30} (\bibinfo {year} {2016})}\BibitemShut
  {NoStop}%
\bibitem [{\citenamefont {Crochik}\ and\ \citenamefont
  {Tom{\'e}}(2005)}]{crochikdakis2005}%
  \BibitemOpen
  \bibfield  {author} {\bibinfo {author} {\bibfnamefont {L.}~\bibnamefont
  {Crochik}}\ and\ \bibinfo {author} {\bibfnamefont {T.}~\bibnamefont
  {Tom{\'e}}},\ }\href@noop {} {\bibfield  {journal} {\bibinfo  {journal}
  {Physical Review E}\ }\textbf {\bibinfo {volume} {72}},\ \bibinfo {pages}
  {057103} (\bibinfo {year} {2005})}\BibitemShut {NoStop}%
\bibitem [{\citenamefont {Stone}\ and\ \citenamefont
  {McKay}(2015)}]{Stone2015437}%
  \BibitemOpen
  \bibfield  {author} {\bibinfo {author} {\bibfnamefont {T.~E.}\ \bibnamefont
  {Stone}}\ and\ \bibinfo {author} {\bibfnamefont {S.~R.}\ \bibnamefont
  {McKay}},\ }\href {\doibase http://dx.doi.org/10.1016/j.physa.2014.10.032}
  {\bibfield  {journal} {\bibinfo  {journal} {Physica A: Statistical Mechanics
  and its Applications}\ }\textbf {\bibinfo {volume} {419}},\ \bibinfo {pages}
  {437 } (\bibinfo {year} {2015})}\BibitemShut {NoStop}%
\bibitem [{\citenamefont {Vilela}\ and\ \citenamefont
  {Moreira}(2009)}]{Vilela20094171}%
  \BibitemOpen
  \bibfield  {author} {\bibinfo {author} {\bibfnamefont {A.~L.~M.}\
  \bibnamefont {Vilela}}\ and\ \bibinfo {author} {\bibfnamefont {F.~G.~B.}\
  \bibnamefont {Moreira}},\ }\href {\doibase
  http://dx.doi.org/10.1016/j.physa.2009.06.046} {\bibfield  {journal}
  {\bibinfo  {journal} {Physica A: Statistical Mechanics and its Applications}\
  }\textbf {\bibinfo {volume} {388}},\ \bibinfo {pages} {4171 } (\bibinfo
  {year} {2009})}\BibitemShut {NoStop}%
\bibitem [{\citenamefont {Stauffer}\ and\ \citenamefont
  {Kulakowski}(2008)}]{1742-5468-2008-04-P04021}%
  \BibitemOpen
  \bibfield  {author} {\bibinfo {author} {\bibfnamefont {D.}~\bibnamefont
  {Stauffer}}\ and\ \bibinfo {author} {\bibfnamefont {K.}~\bibnamefont
  {Kulakowski}},\ }\href {http://stacks.iop.org/1742-5468/2008/i=04/a=P04021}
  {\bibfield  {journal} {\bibinfo  {journal} {Journal of Statistical Mechanics:
  Theory and Experiment}\ }\textbf {\bibinfo {volume} {2008}},\ \bibinfo
  {pages} {P04021} (\bibinfo {year} {2008})}\BibitemShut {NoStop}%
\bibitem [{\citenamefont {Drouffe}\ and\ \citenamefont
  {Godrèche}(1999)}]{0305-4470-32-2-003}%
  \BibitemOpen
  \bibfield  {author} {\bibinfo {author} {\bibfnamefont {J.-M.}\ \bibnamefont
  {Drouffe}}\ and\ \bibinfo {author} {\bibfnamefont {C.}~\bibnamefont
  {Godrèche}},\ }\href {http://stacks.iop.org/0305-4470/32/i=2/a=003}
  {\bibfield  {journal} {\bibinfo  {journal} {Journal of Physics A:
  Mathematical and General}\ }\textbf {\bibinfo {volume} {32}},\ \bibinfo
  {pages} {249} (\bibinfo {year} {1999})}\BibitemShut {NoStop}%
\bibitem [{\citenamefont {Derrida}\ \emph {et~al.}(1991)\citenamefont
  {Derrida}, \citenamefont {Lebowitz}, \citenamefont {Speer},\ and\
  \citenamefont {Spohn}}]{0305-4470-24-20-015}%
  \BibitemOpen
  \bibfield  {author} {\bibinfo {author} {\bibfnamefont {B.}~\bibnamefont
  {Derrida}}, \bibinfo {author} {\bibfnamefont {J.~L.}\ \bibnamefont
  {Lebowitz}}, \bibinfo {author} {\bibfnamefont {E.~R.}\ \bibnamefont {Speer}},
  \ and\ \bibinfo {author} {\bibfnamefont {H.}~\bibnamefont {Spohn}},\ }\href
  {http://stacks.iop.org/0305-4470/24/i=20/a=015} {\bibfield  {journal}
  {\bibinfo  {journal} {Journal of Physics A: Mathematical and General}\
  }\textbf {\bibinfo {volume} {24}},\ \bibinfo {pages} {4805} (\bibinfo {year}
  {1991})}\BibitemShut {NoStop}%
\bibitem [{\citenamefont {Costa}\ and\ \citenamefont {de~Souza}(2005)}]{lsac}%
  \BibitemOpen
  \bibfield  {author} {\bibinfo {author} {\bibfnamefont {L.~S.~A.}\
  \bibnamefont {Costa}}\ and\ \bibinfo {author} {\bibfnamefont {A.~J.~F.}\
  \bibnamefont {de~Souza}},\ }\href@noop {} {\bibfield  {journal} {\bibinfo
  {journal} {Physical Review E}\ }\textbf {\bibinfo {volume} {71}},\ \bibinfo
  {pages} {056124} (\bibinfo {year} {2005})}\BibitemShut {NoStop}%
\bibitem [{\citenamefont {Lima}(2015)}]{lima2014}%
  \BibitemOpen
  \bibfield  {author} {\bibinfo {author} {\bibfnamefont {F.~W.~S.}\
  \bibnamefont {Lima}},\ }\href {\doibase 10.1142/S0129183115500357} {\bibfield
   {journal} {\bibinfo  {journal} {International Journal of Modern Physics C}\
  }\textbf {\bibinfo {volume} {26}},\ \bibinfo {pages} {1550035} (\bibinfo
  {year} {2015})}\BibitemShut {NoStop}%
\bibitem [{\citenamefont {Pereira}\ and\ \citenamefont
  {Moreira}(2005)}]{felipe}%
  \BibitemOpen
  \bibfield  {author} {\bibinfo {author} {\bibfnamefont {L.~F.~C.}\
  \bibnamefont {Pereira}}\ and\ \bibinfo {author} {\bibfnamefont {F.~G.~B.}\
  \bibnamefont {Moreira}},\ }\href {\doibase 10.1103/PhysRevE.71.016123}
  {\bibfield  {journal} {\bibinfo  {journal} {Phys. Rev. E}\ }\textbf {\bibinfo
  {volume} {71}},\ \bibinfo {pages} {016123} (\bibinfo {year}
  {2005})}\BibitemShut {NoStop}%
\bibitem [{\citenamefont {Sampaio-Filho}\ and\ \citenamefont
  {Moreira}(2013)}]{PhysRevE.88.032142}%
  \BibitemOpen
  \bibfield  {author} {\bibinfo {author} {\bibfnamefont {C.~I.~N.}\
  \bibnamefont {Sampaio-Filho}}\ and\ \bibinfo {author} {\bibfnamefont
  {F.~G.~B.}\ \bibnamefont {Moreira}},\ }\href {\doibase
  10.1103/PhysRevE.88.032142} {\bibfield  {journal} {\bibinfo  {journal} {Phys.
  Rev. E}\ }\textbf {\bibinfo {volume} {88}},\ \bibinfo {pages} {032142}
  (\bibinfo {year} {2013})}\BibitemShut {NoStop}%
\bibitem [{\citenamefont {Mendes}\ and\ \citenamefont
  {Santos}(1998)}]{PhysRevE.57.108}%
  \BibitemOpen
  \bibfield  {author} {\bibinfo {author} {\bibfnamefont {J.~F.~F.}\
  \bibnamefont {Mendes}}\ and\ \bibinfo {author} {\bibfnamefont {M.~A.}\
  \bibnamefont {Santos}},\ }\href {\doibase 10.1103/PhysRevE.57.108} {\bibfield
   {journal} {\bibinfo  {journal} {Phys. Rev. E}\ }\textbf {\bibinfo {volume}
  {57}},\ \bibinfo {pages} {108} (\bibinfo {year} {1998})}\BibitemShut
  {NoStop}%
\bibitem [{\citenamefont {Yang}\ \emph {et~al.}(2008)\citenamefont {Yang},
  \citenamefont {Kim},\ and\ \citenamefont {Kwak}}]{YANG}%
  \BibitemOpen
  \bibfield  {author} {\bibinfo {author} {\bibfnamefont {J.~S.}\ \bibnamefont
  {Yang}}, \bibinfo {author} {\bibfnamefont {I.~M.}\ \bibnamefont {Kim}}, \
  and\ \bibinfo {author} {\bibfnamefont {W.}~\bibnamefont {Kwak}},\ }\href@noop
  {} {\bibfield  {journal} {\bibinfo  {journal} {Physical Review E}\ }\textbf
  {\bibinfo {volume} {77}},\ \bibinfo {pages} {051122} (\bibinfo {year}
  {2008})}\BibitemShut {NoStop}%
\bibitem [{\citenamefont {\'Odor}(2004)}]{RevModPhys.76.663}%
  \BibitemOpen
  \bibfield  {author} {\bibinfo {author} {\bibfnamefont {G.}~\bibnamefont
  {\'Odor}},\ }\href {\doibase 10.1103/RevModPhys.76.663} {\bibfield  {journal}
  {\bibinfo  {journal} {Rev. Mod. Phys.}\ }\textbf {\bibinfo {volume} {76}},\
  \bibinfo {pages} {663} (\bibinfo {year} {2004})}\BibitemShut {NoStop}%
\bibitem [{\citenamefont {{Acu\~na-Lara}}\ and\ \citenamefont
  {Sastre}(2012)}]{PhysRevE.86.041123}%
  \BibitemOpen
  \bibfield  {author} {\bibinfo {author} {\bibfnamefont {A.~L.}\ \bibnamefont
  {{Acu\~na-Lara}}}\ and\ \bibinfo {author} {\bibfnamefont {F.}~\bibnamefont
  {Sastre}},\ }\href {\doibase 10.1103/PhysRevE.86.041123} {\bibfield
  {journal} {\bibinfo  {journal} {Phys. Rev. E}\ }\textbf {\bibinfo {volume}
  {86}},\ \bibinfo {pages} {041123} (\bibinfo {year} {2012})}\BibitemShut
  {NoStop}%
\bibitem [{\citenamefont {Grinstein}\ \emph {et~al.}(1985)\citenamefont
  {Grinstein}, \citenamefont {Jayaprakash},\ and\ \citenamefont
  {He}}]{PhysRevLett.55.2527}%
  \BibitemOpen
  \bibfield  {author} {\bibinfo {author} {\bibfnamefont {G.}~\bibnamefont
  {Grinstein}}, \bibinfo {author} {\bibfnamefont {C.}~\bibnamefont
  {Jayaprakash}}, \ and\ \bibinfo {author} {\bibfnamefont {Y.}~\bibnamefont
  {He}},\ }\href {\doibase 10.1103/PhysRevLett.55.2527} {\bibfield  {journal}
  {\bibinfo  {journal} {Phys. Rev. Lett.}\ }\textbf {\bibinfo {volume} {55}},\
  \bibinfo {pages} {2527} (\bibinfo {year} {1985})}\BibitemShut {NoStop}%
\bibitem [{\citenamefont {{NVIDIA Corporation}}(2015)}]{CUDA}%
  \BibitemOpen
  \bibfield  {author} {\bibinfo {author} {\bibnamefont {{NVIDIA
  Corporation}}},\ }\href@noop {} {\emph {\bibinfo {title} {NVIDIA CUDA Compute
  Unified Device Architecture Programming Guide}}}\ (\bibinfo  {publisher}
  {NVIDIA Corporation},\ \bibinfo {year} {2015})\BibitemShut {NoStop}%
\bibitem [{\citenamefont {Preis}\ \emph {et~al.}(2009)\citenamefont {Preis},
  \citenamefont {Virnau}, \citenamefont {Paul},\ and\ \citenamefont
  {Schneider}}]{pvps_jcp_2009}%
  \BibitemOpen
  \bibfield  {author} {\bibinfo {author} {\bibfnamefont {T.}~\bibnamefont
  {Preis}}, \bibinfo {author} {\bibfnamefont {P.}~\bibnamefont {Virnau}},
  \bibinfo {author} {\bibfnamefont {W.}~\bibnamefont {Paul}}, \ and\ \bibinfo
  {author} {\bibfnamefont {J.~J.}\ \bibnamefont {Schneider}},\ }\href@noop {}
  {\bibfield  {journal} {\bibinfo  {journal} {Journal of Computational
  Physics}\ }\textbf {\bibinfo {volume} {228}},\ \bibinfo {pages} {4468}
  (\bibinfo {year} {2009})}\BibitemShut {NoStop}%
\bibitem [{\citenamefont {Oliveira}(1991)}]{murilinho}%
  \BibitemOpen
  \bibfield  {author} {\bibinfo {author} {\bibfnamefont {P.~M.~C.}\
  \bibnamefont {Oliveira}},\ }\href@noop {} {\emph {\bibinfo {title} {Computing
  Boolean Statistical Models}}}\ (\bibinfo  {publisher} {World Scientific},\
  \bibinfo {year} {1991})\BibitemShut {NoStop}%
\bibitem [{\citenamefont {de~Souza}\ and\ \citenamefont
  {Moreira}(1993)}]{PhysRevB.48.9586}%
  \BibitemOpen
  \bibfield  {author} {\bibinfo {author} {\bibfnamefont {A.~J.~F.}\
  \bibnamefont {de~Souza}}\ and\ \bibinfo {author} {\bibfnamefont {F.~G.~B.}\
  \bibnamefont {Moreira}},\ }\href {\doibase 10.1103/PhysRevB.48.9586}
  {\bibfield  {journal} {\bibinfo  {journal} {Phys. Rev. B}\ }\textbf {\bibinfo
  {volume} {48}},\ \bibinfo {pages} {9586} (\bibinfo {year}
  {1993})}\BibitemShut {NoStop}%
\bibitem [{\citenamefont {Menyh\'ard}\ and\ \citenamefont
  {\'Odor}(2000)}]{Odor2000}%
  \BibitemOpen
  \bibfield  {author} {\bibinfo {author} {\bibfnamefont {N.}~\bibnamefont
  {Menyh\'ard}}\ and\ \bibinfo {author} {\bibfnamefont {G.}~\bibnamefont
  {\'Odor}},\ }\href {\doibase
  http://dx.doi.org/10.1590/S0103-97332000000100011} {\bibfield  {journal}
  {\bibinfo  {journal} {Brazilian Journal of Physics}\ }\textbf {\bibinfo
  {volume} {30}},\ \bibinfo {pages} {113 } (\bibinfo {year}
  {2000})}\BibitemShut {NoStop}%
\bibitem [{\citenamefont {Alves}\ \emph {et~al.}(2000)\citenamefont {Alves},
  \citenamefont {de~Felicio},\ and\ \citenamefont {Hansmann}}]{Alves_2000}%
  \BibitemOpen
  \bibfield  {author} {\bibinfo {author} {\bibfnamefont {N.~A.}\ \bibnamefont
  {Alves}}, \bibinfo {author} {\bibfnamefont {J.~R.~D.}\ \bibnamefont
  {de~Felicio}}, \ and\ \bibinfo {author} {\bibfnamefont {U.~H.~E.}\
  \bibnamefont {Hansmann}},\ }\href {\doibase 10.1088/0305-4470/33/42/302}
  {\bibfield  {journal} {\bibinfo  {journal} {Journal of Physics A:
  Mathematical and General}\ }\textbf {\bibinfo {volume} {33}},\ \bibinfo
  {pages} {7489} (\bibinfo {year} {2000})}\BibitemShut {NoStop}%
\bibitem [{\citenamefont {Volpati}\ \emph {et~al.}(2017)\citenamefont
  {Volpati}, \citenamefont {Basu}, \citenamefont {Caracciolo},\ and\
  \citenamefont {Gambassi}}]{PhysRevE.96.052136}%
  \BibitemOpen
  \bibfield  {author} {\bibinfo {author} {\bibfnamefont {V.}~\bibnamefont
  {Volpati}}, \bibinfo {author} {\bibfnamefont {U.}~\bibnamefont {Basu}},
  \bibinfo {author} {\bibfnamefont {S.}~\bibnamefont {Caracciolo}}, \ and\
  \bibinfo {author} {\bibfnamefont {A.}~\bibnamefont {Gambassi}},\ }\href
  {\doibase 10.1103/PhysRevE.96.052136} {\bibfield  {journal} {\bibinfo
  {journal} {Phys. Rev. E}\ }\textbf {\bibinfo {volume} {96}},\ \bibinfo
  {pages} {052136} (\bibinfo {year} {2017})}\BibitemShut {NoStop}%
\bibitem [{\citenamefont {Ferrenberg}\ \emph {et~al.}(2018)\citenamefont
  {Ferrenberg}, \citenamefont {Xu},\ and\ \citenamefont
  {Landau}}]{PhysRevE.97.043301}%
  \BibitemOpen
  \bibfield  {author} {\bibinfo {author} {\bibfnamefont {A.~M.}\ \bibnamefont
  {Ferrenberg}}, \bibinfo {author} {\bibfnamefont {J.}~\bibnamefont {Xu}}, \
  and\ \bibinfo {author} {\bibfnamefont {D.~P.}\ \bibnamefont {Landau}},\
  }\href {\doibase 10.1103/PhysRevE.97.043301} {\bibfield  {journal} {\bibinfo
  {journal} {Phys. Rev. E}\ }\textbf {\bibinfo {volume} {97}},\ \bibinfo
  {pages} {043301} (\bibinfo {year} {2018})}\BibitemShut {NoStop}%
\bibitem [{\citenamefont {Lundow}\ and\ \citenamefont
  {Campbell}(2018)}]{LUNDOW201840}%
  \BibitemOpen
  \bibfield  {author} {\bibinfo {author} {\bibfnamefont {P.}~\bibnamefont
  {Lundow}}\ and\ \bibinfo {author} {\bibfnamefont {I.}~\bibnamefont
  {Campbell}},\ }\href {\doibase https://doi.org/10.1016/j.physa.2018.06.087}
  {\bibfield  {journal} {\bibinfo  {journal} {Physica A: Statistical Mechanics
  and its Applications}\ }\textbf {\bibinfo {volume} {511}},\ \bibinfo {pages}
  {40 } (\bibinfo {year} {2018})}\BibitemShut {NoStop}%
\bibitem [{\citenamefont {Wansleben}\ and\ \citenamefont
  {Landau}(1991)}]{PhysRevB.43.6006}%
  \BibitemOpen
  \bibfield  {author} {\bibinfo {author} {\bibfnamefont {S.}~\bibnamefont
  {Wansleben}}\ and\ \bibinfo {author} {\bibfnamefont {D.~P.}\ \bibnamefont
  {Landau}},\ }\href {\doibase 10.1103/PhysRevB.43.6006} {\bibfield  {journal}
  {\bibinfo  {journal} {Phys. Rev. B}\ }\textbf {\bibinfo {volume} {43}},\
  \bibinfo {pages} {6006} (\bibinfo {year} {1991})}\BibitemShut {NoStop}%
\bibitem [{\citenamefont {Hasenbusch}(2020)}]{PhysRevE.101.022126}%
  \BibitemOpen
  \bibfield  {author} {\bibinfo {author} {\bibfnamefont {M.}~\bibnamefont
  {Hasenbusch}},\ }\href {\doibase 10.1103/PhysRevE.101.022126} {\bibfield
  {journal} {\bibinfo  {journal} {Phys. Rev. E}\ }\textbf {\bibinfo {volume}
  {101}},\ \bibinfo {pages} {022126} (\bibinfo {year} {2020})}\BibitemShut
  {NoStop}%
\bibitem [{\citenamefont {Jaster}\ \emph {et~al.}(1999)\citenamefont {Jaster},
  \citenamefont {Mainville}, \citenamefont {Schülke},\ and\ \citenamefont
  {Zheng}}]{Jaster}%
  \BibitemOpen
  \bibfield  {author} {\bibinfo {author} {\bibfnamefont {A.}~\bibnamefont
  {Jaster}}, \bibinfo {author} {\bibfnamefont {J.}~\bibnamefont {Mainville}},
  \bibinfo {author} {\bibfnamefont {L.}~\bibnamefont {Schülke}}, \ and\
  \bibinfo {author} {\bibfnamefont {B.}~\bibnamefont {Zheng}},\ }\href
  {http://stacks.iop.org/0305-4470/32/i=8/a=008} {\bibfield  {journal}
  {\bibinfo  {journal} {Journal of Physics A: Mathematical and General}\
  }\textbf {\bibinfo {volume} {32}},\ \bibinfo {pages} {1395} (\bibinfo {year}
  {1999})}\BibitemShut {NoStop}%
\end{thebibliography}
%

\end{document}